# Quantum-Accurate Machine Learning Potentials for Metal-Organic Frameworks using Temperature Driven Active Learning

Abhishek Sharma* and Stefano Sanvito*

School of Physics, AMBER and CRANN Institute, Trinity College, Dublin 2, Ireland.

Corresponding author emails: asharma.ms.in@gmail.com (A.S.), sanvitos@tcd.ie (S.S.)

## Abstract

Understanding how structural flexibility affects the properties of metal-organic frameworks (MOFs) is crucial for the design of better MOFs for targeted applications. Flexible MOFs can be studied with molecular dynamics simulations, whose accuracy depends on the force-field used to describe the interatomic interactions. Density functional theory (DFT) and quantum-chemistry methods are highly accurate, but the computational overheads limit their use in long time-dependent simulations for large systems. In contrast, classical force fields usually struggle with the description of coordination bonds.

In this work we develop a DFT-accurate machine-learning spectral neighbor analysis potential, trained on DFT energies, forces and stress tensors, for two representative MOFs, namely ZIF-8 and MOF-5. Their structural and vibrational properties are then studied as a function of temperature and tightly compared with available experimental data. Most importantly, we demonstrate an active-learning algorithm, based on mapping the relevant internal coordinates, which drastically reduces the number of training data to be computed at the DFT level. Thus, the workflow presented here appears as an efficient strategy for the study of flexible MOFs with DFT accuracy, but at a fraction of the DFT computational cost.

# Introduction

Compounds presenting nanometer-size voids form a promising materials platform for various applications, including selective gas diffusion, adsorption and catalysis.[1,2] Flexible metal-organic frameworks (MOFs) have emerged as an intriguing class of nanoporous materials, which allow one to dynamically tune and control the structure and properties of such voids.[3–5] MOFs are crystalline materials made through reticular chemistry, where organic linkers are connected to metal units via coordination bonds. The flexibility of MOFs, in combination with external stimuli, affects the pores and pore channels and gives rise to interesting properties such as linker rotation, gate opening, swelling, negative thermal expansion, negative adsorption etc.[6–9]

In order to study computationally the effects of an external stimulus, such as pressure and temperature, on the properties of flexible MOFs, a detailed analysis of the framework dynamics at an extended length and time scale is necessary.[4,10–13] This can be performed through molecular dynamics (MD) simulations. *Ab-initio* MD (AIMD), as implemented for instance with density functional theory (DFT), provides the most accurate estimation of the potential energy surface (PES). However, the computational overheads are significant and hence AIMD simulations are usually limited to few hundreds of atoms and pico-second time scales. Alternatively, one can use classical interatomic potential models or force-fields. These approximate the PES of a material with the help of parametric functions and may provide estimates of energy, forces, and virial stress of thousand-atom atomic configurations in a short time. However, the use of classical force-fields for MOFs is hampered by their poor performance with atomic environments presenting coordination bonds.[14] Despite this limitation, a variety of classical force-fields have been used to study the properties of MOFs.[15–24] These force-fields are either transferrable (e.g., UFF[15], DREIDING[18], UFF4MOF[16,17] etc.) or developed for a specific MOF (e.g., QUICK-FF[20] and MOF-FF[22,24]).

A possible strategy to achieve DFT accuracy at a computational cost comparable to that of classical force fields is provided by machine-learning potentials (MLPs).[25–31] In these, the atomic chemical environments are represented by mathematical descriptors at various levels of complexity[32], while the corresponding fitting parameters are obtained by training appropriate machine-learning models over the energy, forces, and virial-stress values of a large number of configurations. These are typically obtained by DFT. The accuracy of a MLP to predict the PES of a material depends on the chemical-environment descriptors, the number of parameters in the model, the size and diversity of the training set, and the training procedure.

Recently several computational works have used MLPs to study MOFs.[13,33–42] Most of these employ neural-network potentials (NNPs), which require thousands of training configurations and comprise hundred thousands of parameters to fit. In most of the earlier works, the training configurations were generated via picosecond-long AIMD simulations, which require a long computational time and significant computational resources. Furthermore, the fit of the many parameters is also computationally intensive and the model are little interpretable, namely it is not simple to define from the outset the boundary of the model's validity (e.g., the temperature and pressure range).

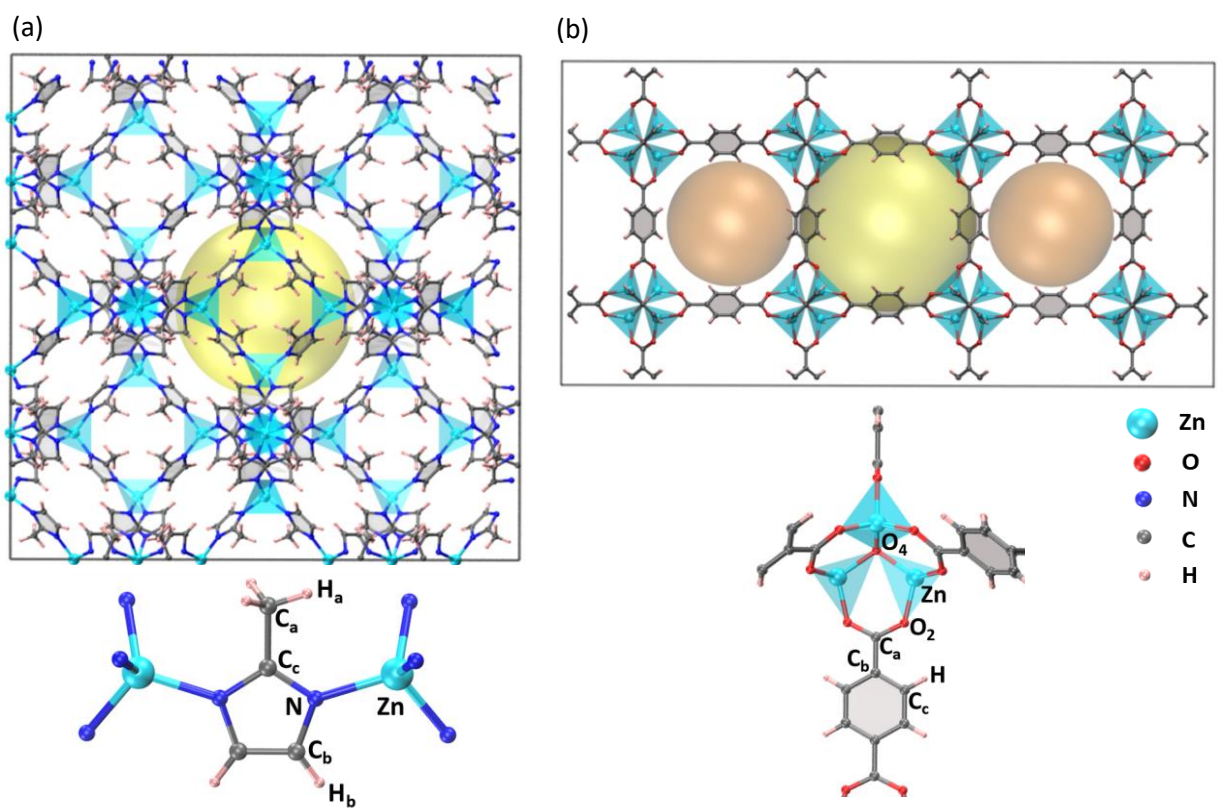

**Figure 1**: Atomic structures of (a) ZIF-8 and (b) MOF-5 (cyan: zinc, blue: nitrogen, red: oxygen, grey: carbon, pink: hydrogen). The basic building units with atom types ($C_a$, $C_b$, etc.) are shown below the main structures. In order to show clearly the pores (yellow and orange spheres), supercells of size 2x2x2 and 1x1x2 are shown for ZIF-8 and MOF-5, respectively. In all calculations described in this work, the unit cell of both ZIF-8 (containing 276 atoms) and MOF-5 (containing 424 atoms) are used.

In this work, we use the spectral neighbor analysis potential (SNAP)[43] as MLP model to study the structural and vibrational properties of MOFs at finite temperature and pressure. SNAP was previously shown to perform well for organic molecules and coordination complexes, and thus it appears as a natural choice for MOFs.[44] SNAP is based on many-body descriptors and linear models, hence, when compared to neural network potentials, it requires only a few hundreds of parameters to obtain similarly accurate fits. For this reason, SNAP typically demands a much smaller training set than those needed by neural networks, so that a limited number of DFT calculations is necessary. As a test bench, here we develop two SNAPs for the widely studied ZIF-8[45] and MOF-5[46] MOFs (their structures are shown in Figure 1). These two particular MOFs have been selected for our study, since various experimental results are available, so that the validity of our approach can be thoroughly tested.

Firstly, we perform a very detailed analysis of the SNAP learning curves, which allows us to propose a protocol for generating correlation-free training sets. Then, we compute various structural and vibrational properties as a function of temperature, and we compare them with available experimental data, demonstrating an excellent agreement. Although applied here to ZIF-8 and MOF-5, our approach is completely general and widely applicable to any other MOFs, whose electronic structure is accessible by DFT (or other *ab initio* electronic structure methods). This allows one to develop predictive SNAPs for MOFs by using only a few hundreds DFT calculations and a simplified training procedure.

## Computational Details and Results

### Active Learning Algorithm

The construction of an adequate training set is crucial for the formulation of a MLP. In fact, MLPs are not physically informed, so that their knowledge of the energy and forces of a particular molecular structure is rooted in having been trained on structures that contain similar local environments. Importantly, in general, MLPs are not guaranteed to extrapolate to poorly known configurations, for which they can catastrophically fail. As such, an ideal training set needs to contain all the local environments that the system will experience when performing the inference, for instance the ones explored during MD simulations. Such training set should also be finely balanced, namely, even when complete, it should not be dominated by a particular pool of local environments. Finally, the size of the training set must be kept as limited as possible, so that the construction of the MLP itself will be numerically convenient in the computational economy of the workflow that one wants to pursue.

With all these requirements in mind we present here an active-learning strategy that allows us to construct a balanced training set, while performing a limited number of DFT calculations. This consists of two main tasks, namely 1) an algorithm that maps the diversity of the training set and ensures that all the relevant local environments are represented, and 2) a strategy to generate the molecular configurations containing those environments. The algorithm is then used for both ZIF-8 and MOF-5, although in two different ways. In fact, although in both cases the first step is identical, then for ZIF-8 the configurations are generated with AIMD, while for MOF-5 they originate from MD runs performed at increasingly large temperatures with subsequently more refined SNAPs. This is because we effectively use the construction of the ZIF-8 SNAP as a learning step to the construction of an efficient method, which is then used for MOF-5.

**Selection of the configurations to include in the training set**

The atomic configuration of a MOF structure can be defined through the knowledge of the unit cell parameters, the atomic bonds, the bond angles, and the dihedral angles [see Figure 2(a)]. Classical force-fields use this information to compute the energy and forces of a configuration in terms of non-bonded and bonded interactions. In the case of non-bonded interactions, atoms are classified into different types, based on their chemical identity and connectivity, information which is then used to define the interaction parameters (e.g., electrostatic charges, Lennard-Jones parameters, etc.). Similarly, bond lengths, bond angles and dihedral angles are classified into different types and correspondingly interaction parameters of the bonded interaction are thus computed.

Inspired by this structure-informed approach, we have developed a simple algorithm to track the diversity and relevance of the local atomic environments included in a training set [see Figure 2(b) for details]. As for the classical force fields, we would like to differentiate structures in terms of a limited number of structural descriptors, namely the cell-parameters, bonds, angles, and dihedrals (collectively called CBAD). The total number of structure descriptors, $N_{CBAD}$, depends on the MOF of interest and consists of 6 cell parameters, $l$ bond types, $m$ angle types, and $n$ dihedral angle types [see Figure 2(a) for details]. Then, we define the resolution of each descriptor, $\Delta$, which is the minimum difference between two values of the descriptor that one can distinguish. This allows us to represent each value of the descriptors as an integer (representing bin index), namely as int($\theta/\Delta$), where $\theta$ is the value of the descriptor. Each MOF configuration has multiple values of each descriptor (except the 6 cell parameters, which have only one value). Therefore, for each MOF configuration we track $N_{CBAD}$ lists of integers, where each list can have multiple integer values. If we wish to consider $M$ possible values for each descriptor in the training set, we will have $M^{N_{CBAD}}$ possible configurations (or configuration matrix) to explore in a $N_{CBAD}$ dimensional configuration space. The question is now how to populate such space with relevant configurations. In principle, one can generate MOF configurations where the descriptors are varied one at a time, but this will necessitate $M^{N_{CBAD}}$ DFT calculations, a number that can be prohibitively large. The strategy chosen here, instead, is that of mapping the entire space with a limited number of configurations.

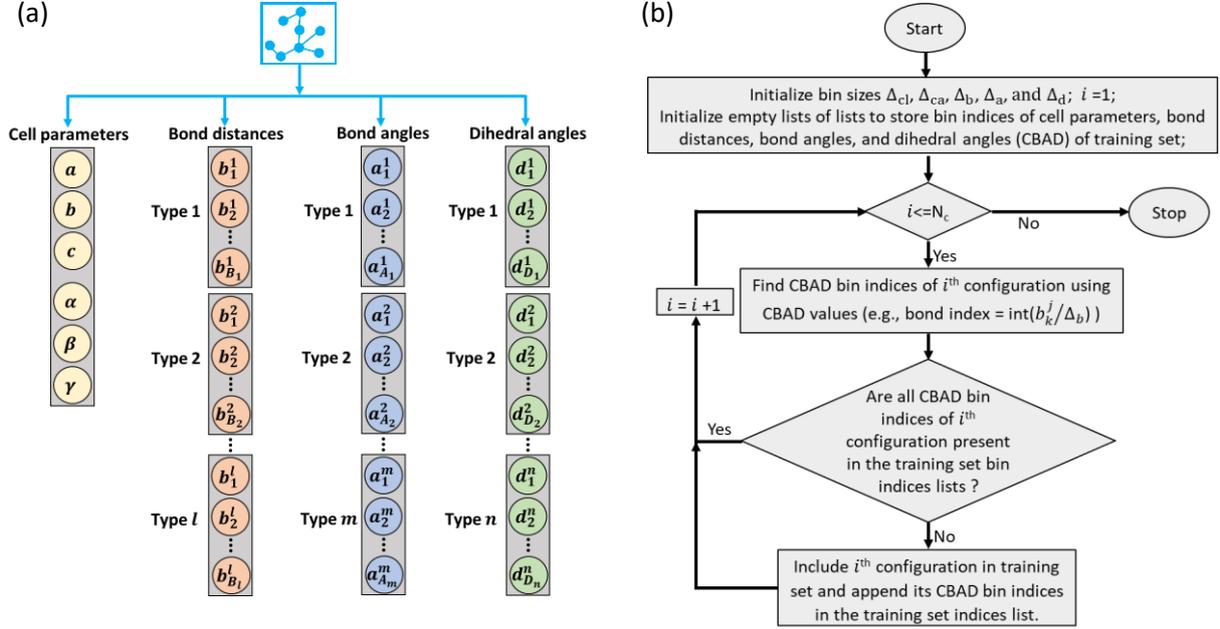

**Figure 2**: (a) Schematic showing the cell parameters (lattice parameters: $a$, $b$, $c$; cell angles: $\alpha$, $\beta$, $\gamma$) and different types of bonds, angles, and dihedrals present in the atomic configuration of a MOF structure. This representation of an atomic configuration is termed here as CBAD (cell parameters, bonds, angles, and dihedrals). (b) A flowchart illustrating our CBAD-based algorithm to select training set configurations from a given set of $N_c$ configurations. Here $\Delta_{cl}$, $\Delta_{ca}$, $\Delta_b$, $\Delta_a$, and $\Delta_d$ are the descriptor resolutions for the cell lengths ($a$, $b$, $c$), cell angles ($\alpha$, $\beta$, $\gamma$), bond distances ($b_k^j$), angles ($a_k^j$), and dihedral angles ($d_k^j$), respectively. The $\Delta_i$ values are used to find CBAD bin indices (e.g., bond index = int($b_k^j/\Delta_b$)). The algorithm then compares the CBAD indices of a new configuration with the CBAD indices of the training set and will include the new configuration in training set, if at least one of the indices is not present in the configuration matrix.

In order to reduce the number of DFT calculations, we consider the multiplicity of descriptor values in an atomic configuration of MOF (within the $N_{CBAD}$ lists of integers). If an atomic configuration results from a MD simulation (at certain temperature), then all values of a descriptor in that configuration would be different and belong to a distribution (of descriptor at certain temperature). In such configuration, the atoms in that configuration can have multiple local chemical environments. Thus, instead of tracking the $M^{N_{CBAD}}$ configuration matrix, we individually track $N_{CBAD}$ descriptors and corresponding configuration matrix of size of $N_{CBAD} \times M$. With this simplification, our algorithm proceeds as follows. An initial configuration defines multiple value for each of the $N_\text{CBAD}$ descriptors, we consider only unique values for each descriptor and populate the values in the $N_\text{CBAD}$ descriptor lists (to sample the $N_\text{CBAD} \times M$ configuration matrix). The next configuration will then define a new $N_\text{CBAD}$ list of descriptors. If at least one of them is not already present in the corresponding descriptor lists, then such configuration will be accepted and it will be part of the training set. In this way we ensure that all the relevant values of each descriptor, within a precision $\Delta$, are represented at least once in our training set. A schematic illustration of the algorithm is provided in Figure 2(b), while a detailed pseudocode is given in Figure S1 of the Supplementary Information (SI).

## Training the SNAP

Within the SNAP formalism[43] the total energy, $E$, of a molecule (or a solid) is expressed as the sum of individual atomic energies, $E_i$. These are, in turn, function of the local chemical environment of each individual atom, which is defined within a radial cut-off. SNAP then expands the local atomic-density distribution over four-dimensional spherical harmonics and constructs the associated bispectrum components, $B_j$, which form a rotationally invariant set of descriptors. Finally, the atomic energies are taken as a linear function of the bispectrum components, namely

$$E = \sum_i^{N_a} E_i = \sum_i^{N_a} \sum_l^{N_{2J}} \beta_i^l B_i^l$$

where, $N_a$ is total number of atoms, $N_{2J}$ is total number of bispectrum functions ($2J$ controls the order of the expansion), and $\beta_i^l$ are the coefficients of the bispectrum components. The accuracy of the chemical-environment description can be tuned by tuning the number of bispectrum components. Earlier work[44] has shown that considering 56 bispectrum components (corresponding to $2J=8$) per chemical species results in a reasonable accuracy, and thus we have used same value here. In both ZIF-8 and MOF-5 there are 7 different atom types (see Figure 1 – note that C, O, and H atoms with different coordination are considered as different atom types), so that our SNAP models are constructed over 392 bispectrum functions (and 392 $\beta_i^l$ values). The SNAP training, namely the computation of the $\beta_i^l$ values, is here performed over the energy, forces and the virial stress of each of the configurations contained in the training set, with the reference values being computed with DFT (see section S1.1 in SI for details). Thus, each configuration provides $3N_a+7$ training data (1 energy, $3N_a$ forces, and 6 virial stress components). The unit cells of ZIF-8 and MOF-5 contain 276 and 424 atoms, respectively. Therefore, if the training set comprises in the region of 600 configurations, we will have approximately 0.5 and 0.7 million of training data for ZIF-8 and MOF-5, respectively. We generate the bispectrum components of a given configuration by using the Large-scale Atomic/Molecular Massively Parallel Simulator (LAMMPS)[47] package and then obtain the bispectrum coefficients through ridge regression. We optimize the SNAP hyperparameters (atomic-species-dependent cutoff radius and chemical-species weights) by using the Scipy package and we drive the optimization by minimizing the error over energy, forces and stress tensor. Then, the SNAP training is performed at the optimal hyperparameters and the final model is used to perform molecular dynamics simulations.

## Generation of the training and test sets for ZIF-8

In order to establish a general SNAP-training protocol for MOFs, we have first developed the potential for ZIF-8. The unit cell of ZIF-8 contains four elements (C, H, N, and Zn), 7 atom types [see Figure 1(a)] and 276 atoms. The different configurations to be included in the training set are generated by first following the computationally intensive approach used in earlier works, namely we perform AIMD simulations (details are given in section S1.1 of the SI). These have a duration of 1 ps (with a 0.5 fs timestep) at temperatures ranging from 100K to 1000K with a 100K interval. From the generated 20,000 configurations, we then select those to include in the training set by using the simple algorithm described in the previous section. Note that for all configurations included in the training and test set, we run high-quality DFT calculations with the higher cut-off of 1000 Ry (details of all DFT calculations are given in section S1.1 of SI).

In general, the number of configurations contained in the training set should be optimal, since a few configurations will result in poor a representation of the PES, while too many configurations are associated to a high computational cost and to a possible imbalance in the representation of the main structural characteristics of the MOF. Thus, finding the optimal number of configurations is an essential step for the development of the MLP. Our strategy to populate the configuration matrix mitigates the risk of oversampling, and we can systematically change the number of configurations by changing the resolution of the structural descriptors (how finely we sample each descriptor). In this way we create training sets ranging from 50 to 3000 configurations and fit a SNAP for each of these training sets. Then, the performance test is conducted over two different sets. The first, referred here as test set A, contains around 5000 configurations obtained from AIMD simulations (at different temperatures between 100 to 1000

K), while the second (test set B) is generated by using classical MD simulations (using force-field proposed by Weng et al.[23]) of the ZIF-8 unit cell at 500 K. In this second case we select around 2000 configurations.

(a) Learning curves for ZIF-8

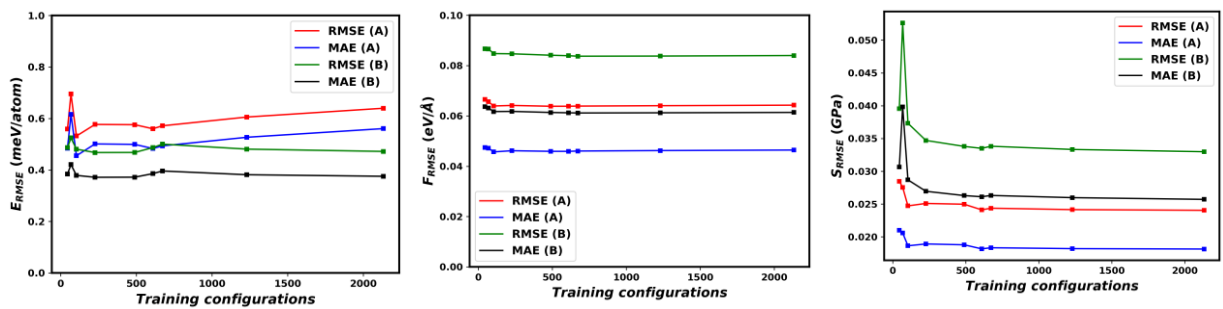

(b) Parity plots for training data of ZIF-8

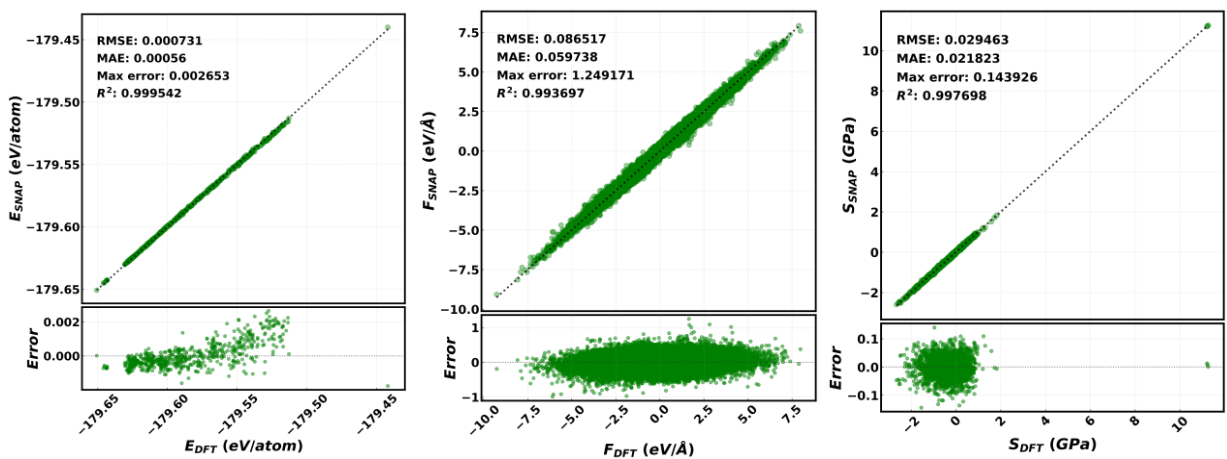

**Figure 3**: SNAP model for ZIF-8. (a) Learning curves for the RMSE and MAE for energy (left-hand side panel), forces (middle panel) and virial stress (right-hand side panel). Data are presented for test set A (composed of ~5000 configurations from AIMD simulations) and test set B (composed of ~2000 configurations from classical MD simulations) as a function of the number of configurations in the training set. Note that no significant change in the error is observed after the training set size reaches ~ 600 configurations. (b) Parity plots for energy, forces and virial stress values comparing DFT and SNAP values. The RMSE is 0.7 meV/atom, 86 meV/Å, and 29.5 MPa, respectively for energy, forces and virial stress.

The learning curves and the associated parity plots, computed over energy, forces and viral stress, are displayed in Figure 3. The learning curves are taken over the test set and display both the root mean square error (RMSE) and the mean absolute error (MAE) as a function of the number of configurations in the training set. Although there are some small subtleties depending on the test set, it is quite clear that the errors plateau for training sets containing in excess of 600 configurations. In fact, for the energy and test set A we observe a marginal error enhancement when the configurations are increased beyond 600, a feature that may suggest some minor overfitting. In any case, the errors of the converged SNAP are extremely low, namely of the order of 0.5 meV/atom, 50 meV/Å and 25 MPa, respectively for energy, forces and stress tensor. This level of accuracy is certainly enough to perform reliable MD over a broad temperature range, as we will demonstrate later on.

**Training and Test Set for MOF-5**

In the construction of the ZIF-8 SNAP we did generate about 20,000 configurations, but then realised that 600 are ideal to fit a high-performing model. Now, for MOF-5 we wish to establish a method that allows us to compute only the 600 configurations needed without any redundancy. In a recent work, an incremental learning approach was used in combination with metadynamics to generate the training set configurations of a MLP.[38] In a metadynamics simulation, a few collective variables are defined and bias is added along their trajectories to explore a particular region of the phase space. Since, increasing the temperature corresponds to enlarging the phase space explored for all structural descriptors (and not just the collective variables), here we develop a simple algorithm driven by temperature to generate the training set for MOFs (see Figure 4 for details).

The proposed algorithm proceeds as following (see Figure 4). Firstly, we take the experimental crystal structure and generate different configurations by introducing a small random perturbation to the atomic positions (Step 0 in Fig. 4). Among these configurations we select those to populate the defined configuration matrix according to the algorithm described before, and their electronic structure is computed by DFT. The DFT energies, forces and stress tensors are then used to train an initial SNAP (MLP$_0$). The following step (Step 1 in Fig. 4) performs 50 independent MD runs, starting from 50 inequivalent configurations obtained by random displacing atoms from the experimental structure. The MD is conducted, starting from different initial velocities, at the low temperature of 100 K (in the *NPT* ensemble) by using MLP$_0$ for approximately 2000 steps. We then select at most one configuration from each MD run to be included in the training set, according to the selection criterion discussed before, and for these we run again DFT simulations. Such expanded training set is then used to construct the next generation of SNAP (MLP$_1$). Step 1 is then repeated multiple times at a progressively higher MD temperature, which is here increased by 100K at each step. This process enhances the diversity of the training set and expands the range of temperature at which the SNAP can be used. For MOF-5 we performed iterations until the temperature reached 1000 K, obtaining a total of 487 training configurations.

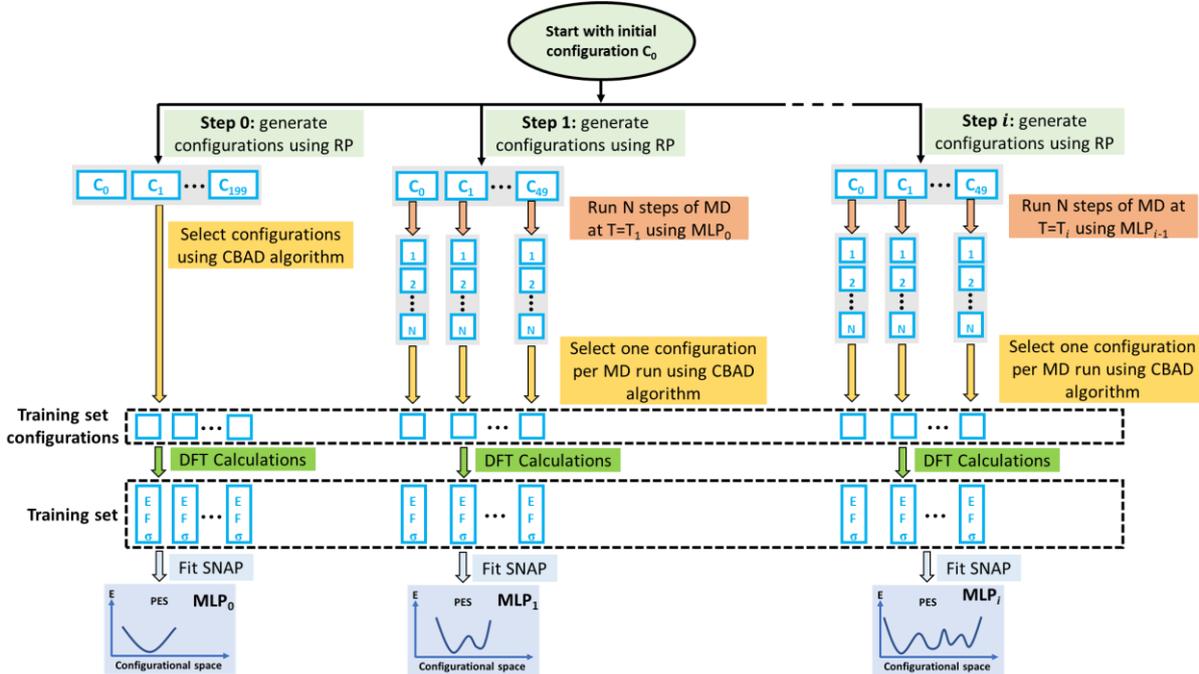

**Figure 4**: Overview of the workflow used to generate the training set for MOF-5. At each step a new SNAP is constructed by using the structure configurations obtained from the MD simulations performed with the SNAP trained at the previous step (see details in the text). 'RP' means random perturbation of the atomic positions.

Once the SNAP corresponding to the highest temperature is constructed, we perform MD simulations with temperature now ramping between 10K to 1000K and select the configurations to be used for the test set (1191 in total). Furthermore, we randomly select additional ~100 configurations to be included in the training set, so that the total number remains close to 600 (596 in our case). The final SNAP is then trained on such training set, and its parity plots are presented in Figure 5. To further validate the trained SNAP, we generate configurations using MD simulations (with trained SNAP) at 300 K. Parity plots for test set are also shown in Figure 5. Again, we obtain a very high-quality potential with training-set energies accurate to sub meV/atom, and MAE on forces and stress components of 100 meV/Å and 19 MPa, respectively. Note that the errors on the test set are even lower than those on the training data. This is due to the fact that the test configurations are generated via MD simulations with a properly trained SNAP (where the temperature is ramped between 10K to 1000K), and therefore, they are less distorted when compared to the training configurations. As a consequence, the range of values for energy, forces and viral stresses in the test set is more limited than that of the training set.

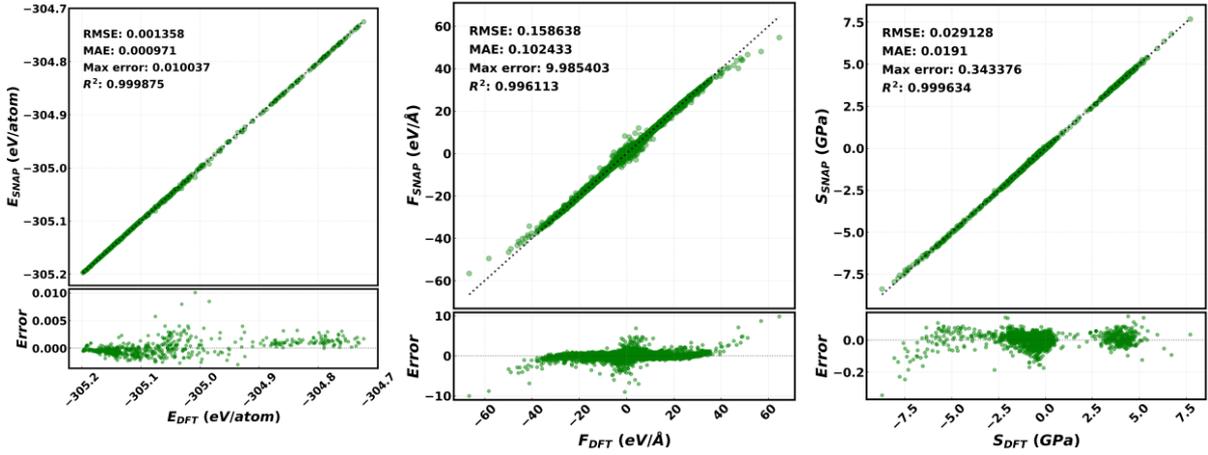

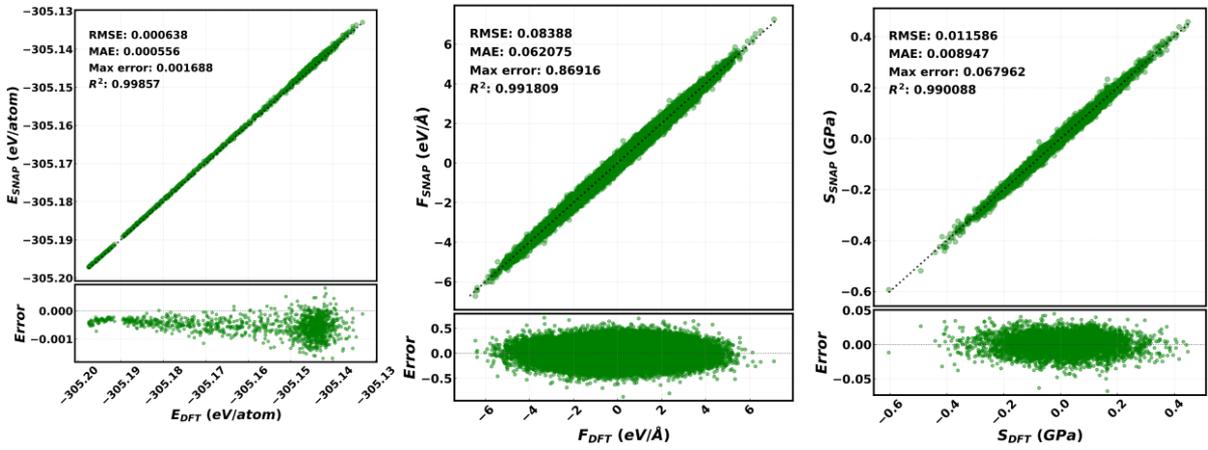

**Figure 5**: Parity plots for energy (left-hand side panel), forces (middle panel) and virial-stress components (right-hand side panel) of the MOF-5 SNAP, computed over the training (596 configurations – upper panel) and test set (1191 configurations – lower panels). The RMSE for the test set are 0.6 meV/atom, 84 meV/Å and 11.6 MPa respectively for energy, forces and virial stress components.

## Molecular Dynamics (MD) Simulations

We now proceed to evaluate a number of structural and vibrational properties ZIF-8 and MOF-5, as a function of temperature. In particular, we begin by looking at the effect of temperature and pressure on the lattice constant. For this we perform molecular dynamics simulations in the isothermal-isobaric ensemble, with constant number of particles, $N$, constant pressure, $P$, and temperature, $T$. All our results are obtained with LAMMPS, and the SNAPs developed in the previous sections (see details of the potentials in section S1.2 of the SI). Note that, in order to account for van der Waals interaction, we have added to the SNAP energies and forces contributions from the Grimm's D3 scheme.[48]

# Performance of Trained SNAP

## Lattice Constant

The trained SNAPs are used to perform five independent 600ps-long MD simulations (with a timestep of 1 fs) in the *NPT* ensemble for both ZIF-8 and MOF-5. Then, the trajectories of the first 100ps are considered as equilibration steps and the remaining 500ps are used for property calculations. A window of 10ps is used to estimate the average and the variance in the lattice parameters, with our results being summarised in Figure 6. In general, we find an excellent agreement between the simulated lattice parameters and available experimental data. In the case of ZIF-8 the lattice parameter increases with temperature (this is referred to as positive thermal expansion) and decreases with pressure. In contrast, MOF-5 has a negative thermal expansion. The computed linear thermal expansion coefficient at 300 K for ZIF-8 is $7.1 \times 10^{-6}$ K$^{-1}$, which is within the experimental range determined by $11.9 \times 10^{-6}$ K$^{-1}$ (Sapnik et. al.[49]) and $6.5 \times 10^{-6}$ K$^{-1}$ (Burtch et. al.[50,51]). Similarly, we obtain a linear thermal expansion coefficient of $-13.3 \times 10^{-6}$ K$^{-1}$ for MOF-5 at 300 K, which is close to experimental value[52] of $-13.1 \times 10^{-6}$ K$^{-1}$. Such excellent agreement indicates that our SNAPs are well capable of describing volumetric changes of the lattice parameters as a function of temperature. Note that the absolute value of the lattice parameters predicted by SNAP is slightly larger than that measured experimentally, by approximately 0.5% for both MOFs. This minor overestimation is due to the use of the DFT generalized-gradient approximation (GGA) to the exchange and correlation functional used for the construction of the training set. GGA sometime may slightly underbind and this feature is here transferred to the SNAP.

## Vibrational Density of States (VDOS)

Having investigated the temperature and pressure response of the MOFs we now move at analysing their vibrational properties. In particular, we compute the vibrational density of states (VDOS), which is here obtained as the Fourier transform of the mass-averaged velocity autocorrelation function along an MD trajectory. In this case, we perform an *NPT* simulation at 300 K for 1ps, followed by a 500ps-long *NVE* simulation, from which we extract atomic configurations and velocities every 2 fs. Our computed VDOSs are shown in Figure 7, while the partial VDOS (PVDOS) projected over each atom type are shown in Figures S4 and S6 of the SI. Similar to results of Eckhoff et al.[33] for MOF-5, here we observe two spectral regions (below 1700 cm$^{-1}$ and after 2900 cm$^{-1}$) for both MOF-5 and ZIF-8. Then, we compare the PVDOS (Figure S5 and S7) of different atoms to identify the modes associated to the various peaks of the vibrational spectrum.

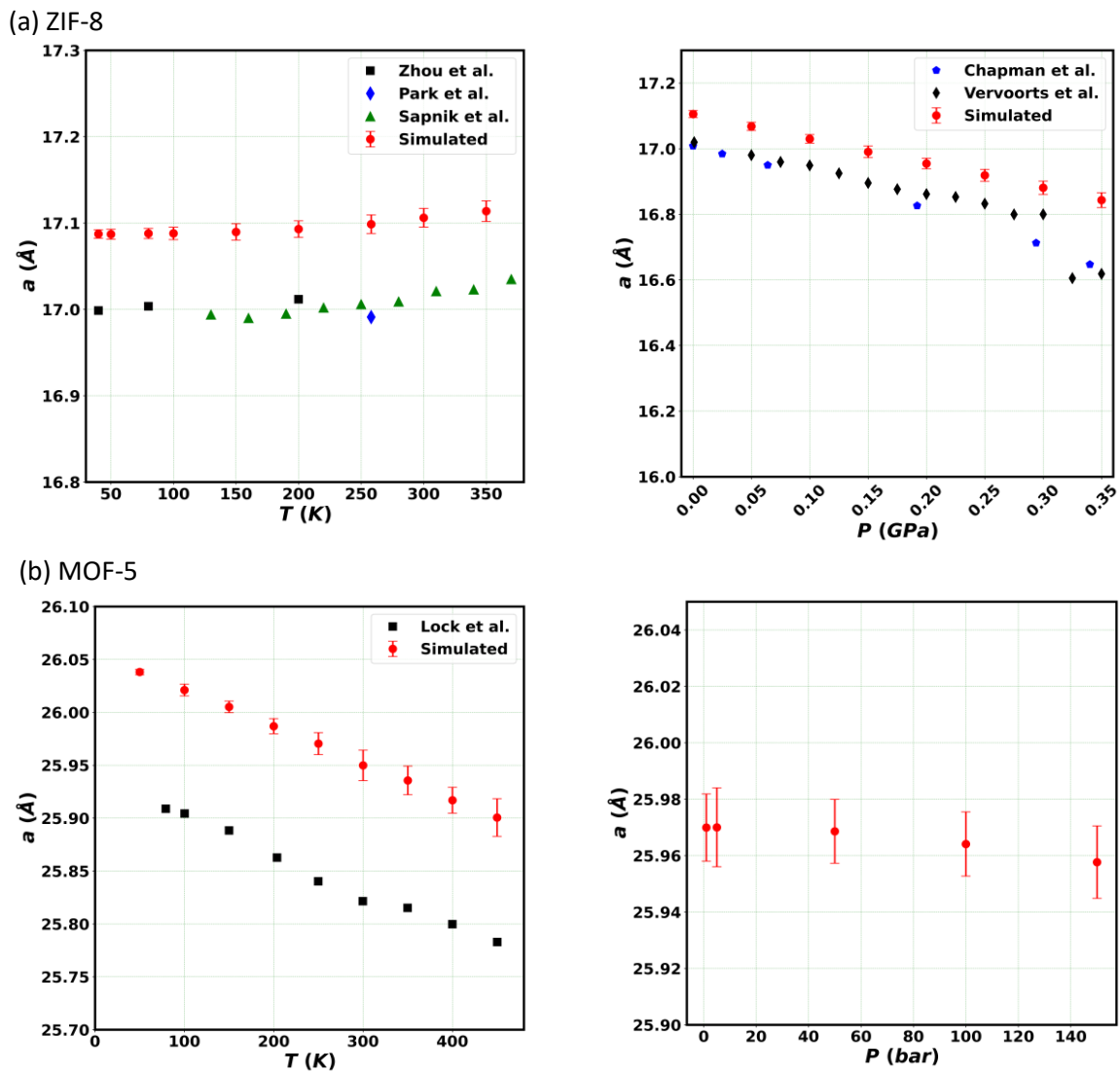

**Figure 6**: Simulated unit cell parameter for (a) ZIF-8 and (b) MOF-5 as a function of temperature (left-hand side panels) and pressure (right-hand side panels), and their comparison with available experimental data (ZIF-8: Zhou et al.,[53] Park et al.,[45] Sapnik et. al.[49], Chapman et al.[54], and Vervoorts et al.[55]; MOF-5: Lock et al.[52]). For ZIF-8 (MOF-5), the pressure is kept fixed at 1 bar during the temperature scan, while the temperature is kept fixed at 300 K (250 K) during the pressure scan.

In general, the experimental[56] infrared (IR) and Raman spectrum of both ZIF-8 and MOF-5 agrees well with our simulated VDOS. For ZIF-8, the PVDOS (Figure S5) reveals Zn-N vibrational frequencies at around 180 cm$^{-1}$, 226 cm$^{-1}$ and 286 cm$^{-1}$, which are close to the experimental Zn-N Raman frequencies[57] of 168 and 273 cm$^{-1}$ and the far infrared (IR) frequencies[58] in the 265-325 cm$^{-1}$ range. We also observe PVDOS peak at 1507 cm$^{-1}$ for $C_b$, $C_c$, and N (see Figure 1 for the definition of the atom types), which are associated to the stretching of the N-$C_b$ and N-$C_c$ bonds. This value is close to experimental Raman frequencies[57] at 1499 and 1508 cm$^{-1}$. For the aromatic $C_b$-$H_b$ dynamics, we observe VDOS spectral amplitude in the 3200-3250 cm$^{-1}$ range, which is close to the Raman frequencies[57] of 3110 and 3131 cm$^{-1}$ and the IR frequency[59] of 3135 cm$^{-1}$. The simulated VDOS for methyl $C_a$-$H_a$ dynamics are observed in the range of 2900-3150 cm$^{-1}$ which is close to Raman frequencies[57] of 2915 and 2931 cm$^{-1}$ and IR frequency[59] of 2927 and 2961 cm$^{-1}$.

Moving to MOF-5, we observe phonon bands up to 1650 cm$^{-1}$ in the first spectral region. The PVDOS (Figure S7) reveals vibrational frequencies of 486-493 cm$^{-1}$, and 556 cm$^{-1}$ for $O_4$-Zn and frequencies of 263 cm$^{-1}$, 363 cm$^{-1}$, and 426-443 cm$^{-1}$ for $O_2$-Zn. These are close to the experimental[56,60] Zn-O IR frequency of 523 cm$^{-1}$. We also observe a good agreement between the experimental[60] C-O vibrational frequency of 1377 and 1585 cm$^{-1}$ with the peaks in the simulated PVDOS.

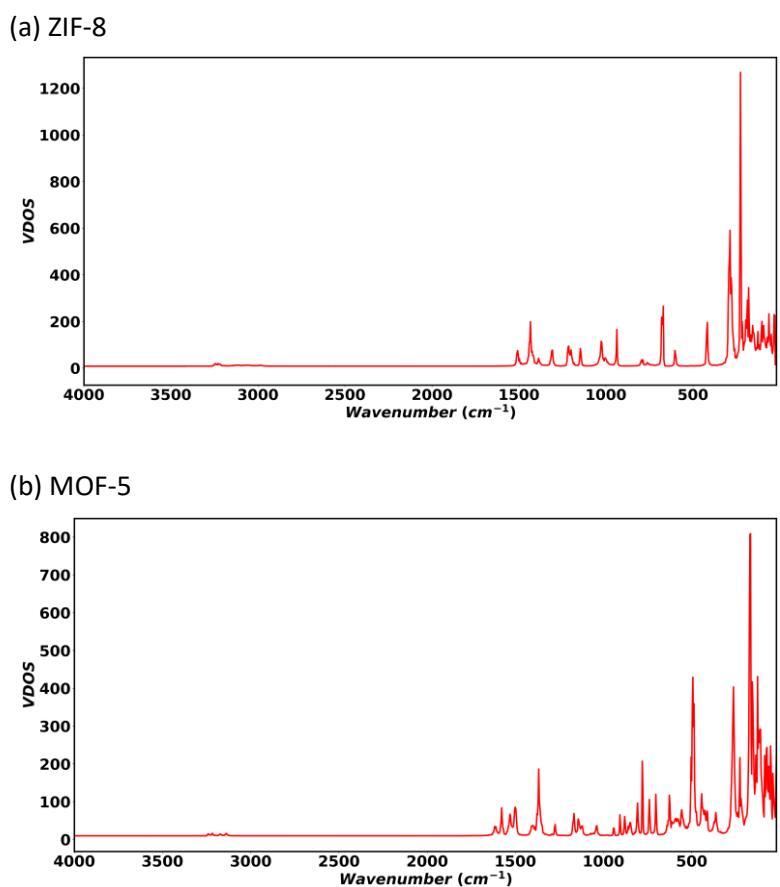

**Figure 7**: Simulated vibrational density of states (VDOS) for (a) ZIF-8 and (b) MOF-5. The corresponding partial VDOSs for each atom types are shown in Fig S4-S7 of the SI.

**Free energy barrier for rotation of the MOF-5 phenylene rings**

In order to unravel the internal dynamics of a MOF, it is essential to develop an understanding of the free energy barriers for the internal dynamics of different groups.[4,61–64] Such barriers can be studied with our MLP, which should be able to reliably map the atomic environments in the transition-state zone of the phase space. In MOF-5 the phenylene rings do not have a significant steric hinderance, however, significant interaction with the neighbouring atoms creates a barrier to their rotational dynamics along the central axis. Therefore, in our MD simulations for MOF-5 at room temperature we did not observed rotation of any phenylene ring.

We have then analysed the distribution of the dihedral angles in MOF-5 (see Figure S8 of SI) in the training set configurations (generated with our temperature-driven active leaning algorithm at temperatures comprised between 100-1000 K). We have found that the training set contains configurations with all possible values of the $C_c$-$C_b$-$C_a$-$O_2$ dihedral angles (-180° to 180°). Thus, with our approach, the trained SNAP can reliably map atomic environments near the transition state corresponding to the rotational barrier. This motivates us to quantify the free-energy barrier for phenylene ring rotation in MOF-5. For this purpose, we perform well-tempered metadynamics (WTM) simulations, considering the two dihedral angles ($\phi_1$ and $\phi_2$) at each side of the phenylene ring as collective variables (see Figure 8). All the WTM simulations are performed in the *NVT* ensemble, therefore, the resultant free energy corresponds to the Helmholtz free energy. Additional details about the WTM simulation are given in section S1.3 of SI.

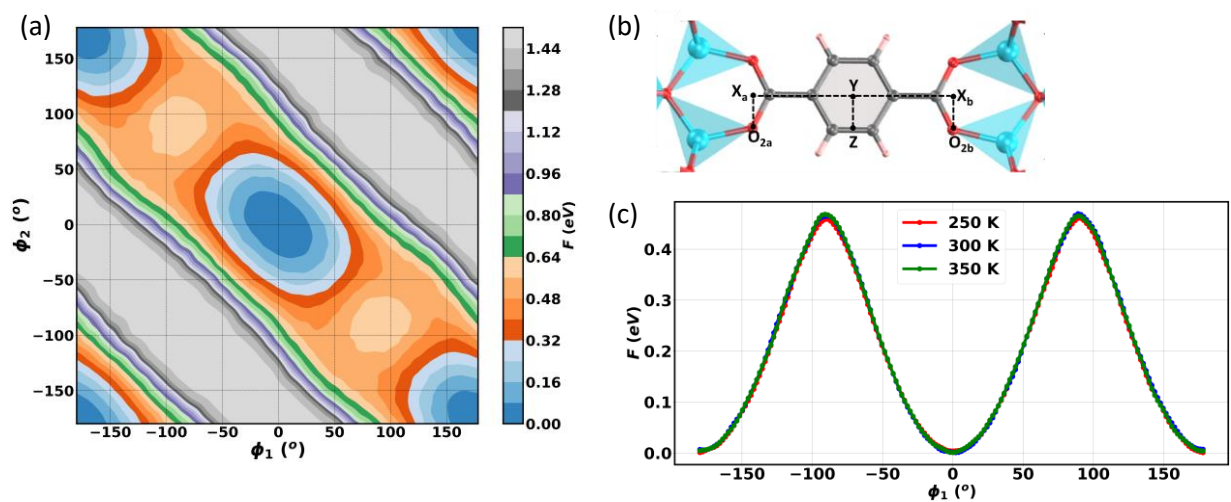

**Figure 8**: (a) Free energy profile for rotation of a benzene ring in MOF-5 as a function of the two dihedral angle collective variables $\phi_1$ and $\phi_2$. The free energy minimum is shown in blue, and the barriers is shown in orange. (b) Schematic showing the collective variables, $\phi_1 = \phi_{O_{2a}X_aYZ}$ and $\phi_2 = \phi_{O_{2b}X_bYZ}$. (c) Free energy for the benzene ring rotation as a function of one of the collective variables at different temperatures. A barrier of 0.46 eV is observed.

The free energy profile as a function of both collective variables is shown in Figure 8(a). The stable states in the rotation (blue regions) are separated by transition states (orange regions). If we consider the simplified case where the MOF-5 structure is rigid and only the rotational motion of the phenylene rings is allowed, one will have a negative correlation between the two collective variables and the only free-energy variation will be on the diagonal line of the 2D mesh of Figure 8(a). Along this diagonal, we compute a free energy barrier of more than 0.6 eV. Since the MOF-5 structure is flexible, the oxygen atoms wobble with respect to the Zn ones, a feature that relaxes the negative correlation between the collective variables and makes the free energy profile broader. This wobbling results in a transition path presenting a lower free-energy barrier than that along diagonal path, hence the wobbling of oxygen atoms helps in the rotation of phenylene rings of MOF-5. The free-energy profile as a function of one dihedral angle, Figure 8(c), is finally obtained by integrating the effect of other angle[4] and a rotation free-energy barrier of 0.46 eV is thus computed. This value is close to the experimental one[64] of 0.49 eV, indicating once again the excellent quality of our interatomic potential.

## Conclusion

In the last few years, the application of machine learning potentials to the study of MOFs has received a growing attention. In this work we have presented a strategy to develop accurate spectral neighbor analysis potentials (SNAPs), a class of machine force fields, for ZIF-8 and MOF-5, two of the most experimentally studied MOFs. SNAP uses many-body structural descriptors together with linear regression, so that it necessitates only a few hundreds parameters and can be trained in a few minutes over a laptop. This computational overhead is much more limited than that requited by neural network potentials, which typically need to optimize more than $10^5$ parameters on a multi-core computational platform. Here we discuss a general and simple scheme to generate a tight and comprehensive training set and propose a simple active-learning algorithm based on populating an appropriate distribution of bond distances, bond angles, dihedral angles, and cell parameters. This is implemented with a hierarchical molecular-dynamics workflow, which uses only SNAP to produce the optimum number of correlation-free training-set configurations. Our approach has then been used to train two SNAPs for ZIF-8 and MOF-5, and hence to the study of their temperature-dependent structural and vibrational properties. In general, we obtain an excellent agreement with experimental data and we are even able to explore the phase-space of MOFs close to the transition state relevant for structural changes. Although applied here only to two examples, our method is general and can be widely deployed to construct high-performing machine-learning potentials at a low computational cost for studying room temperature (and above) MOFs dynamics. Further investigation is necessary to explore the robustness of our approach to the study of MOFs at extreme temperature and pressure conditions and other MOFs with multiple phases.

## Acknowledgements

Authors thank the Trinity Centre for High Performance Computing (TCHPC) and the Irish Centre for High-End Computing (ICHEC) for providing the computational resources. A.S. thank Advanced Materials and BioEngineering Research (AMBER) Centre and science foundation of Ireland (SFI) for providing research fellowship.

# Supporting Information

# Quantum-Accurate Machine Learning Potentials for Metal-Organic Frameworks using Temperature Driven Active Learning

Abhishek Sharma* and Stefano Sanvito*

School of Physics, AMBER and CRANN Institute, Trinity College, Dublin 2, Ireland.

Corresponding author emails: asharma.ms.in@gmail.com (A.S.), sanvitos@tcd.ie (S.S.)

# Section S1: Computational Details

## S1.1 Density Functional Theory (DFT) Calculations

The QUICKSTEP[1] module of CP2K[2] package was used for all DFT calculations. In this approach, Kohn-Sham molecular orbitals are described as a linear combination of atom centred Gaussian-type orbitals. All atoms were described using MOLOPT basis set in combination with norm-conserving Goedecker-Teter-Hutter[3] (GTH) pseudopotentials. The PBE[4] exchange-correlation functional was used and electron density was described as an auxiliary plane-wave basis set with certain energy cut-off (different for different type of calculations). The orbital transformation approach was used to find solution of Kohn-Sham equations. Self-consistent field (SCF) convergence of both outer and inner loops was achieved with an accuracy of $10^{-7}$.

### S1.1.1 Structure Optimization

CP2K was used to perform geometry and cell optimization of both ZIF-8 and MOF-5 structures. In geometry optimization calculations, dispersion corrections were included using the DFT-D3[5] approach with BJ damping and the energy cut-off of 1000 Ry was used.

### S1.1.2 Ab-initio Molecular Dynamics Simulations (AIMD) of ZIF-8

To generate different atomic configurations of ZIF-8 AIMD simulations were performed. We first optimized the unit cell of ZIF-8 (with 276 atoms) using density functional theory (DFT) as implemented in CP2K package (energy cut-off of 600 Ry). We further used CP2K to perform AIMD simulations at different temperatures (100 to 1000 K with gap of 100 K). The MD simulations were performed at constant temperature and pressure using a flexible cell. A timestep of 0.5 fs was used and AIMD simulations were performed for 2000 steps (1 ps) at each temperature. To maintain the temperature, Nose-Hoover thermostat was used with time constant of 50. Pressure was kept at 1 bar with the help of barostat (implemented in CP2K) with time constant of 100.

### S1.1.3 Calculation of Energy, Forces, and Stress Tensor

DFT was used for calculation of energy, forces, and stress tensor values for a given atomic configuration. In these calculations, the energy cut-off of 1000 Ry was used.

## S1.2 Molecular Dynamics Simulation

Trained spectral neighbor analysis potential (SNAP)[6] was used to perform molecular dynamics (MD) simulations of both ZIF-8 and MOF-5 using Large-scale Atomic/Molecular Massively Parallel Simulator (LAMMPS)[7] package. In addition to SNAP, dispersion corrections were also included in MD simulation using the Grimm's D3[5] approach with BJ damping. At each step of MD, the EFS values of D3 correction were added to EFS values of SNAP, and positions and cell lengths were updated accordingly. A timestep of 1 fs was used in all MD simulations.

For ZIF-8, additional repulsive Ziegler-Biersack-Littmark (ZBL)[8] empirical potential was employed to create repulsion between $H_b$-$H_a$, $H_b$-N, $H_b$-Zn, $H_a$-$C_b$, $H_a$-N, and $H_a$-Zn atomic pairs. In ZBL the inner and outer cut-off of 1.8 and 2.3 Å were chosen, respectively. In training set configurations of ZIF-8, the considered atomic pairs for ZBL repulsion have distance more than 2.5 Å, resulting in zero contribution of ZBL in energy, forces, and stress values. Therefore, the effect of ZBL was not subtracted from training set data of energy, forces, and stress values. For MOF-5, no ZBL repulsion was considered.

### S1.2.1 MD Simulations in NPT Ensemble

To determine the performance of SNAP on unit cell lengths, we performed MD simulation in isothermal-isobaric ensemble (NPT: at constant number of particles (N), pressure (P), and temperature (T)). Before performing MD simulations in NPT ensemble, the structures were optimized. To maintain the temperature, Nose-Hoover thermostat with 5 chains was used. To maintain pressure, Nose-Hoover barostat was used. During these simulations, only cell lengths were allowed to change and cell angles were kept constant.

## S1.3 Well-Tempered Metadynamics Simulations

We performed well-tempered metadynamics (WTM)[13,14] simulation using the open-source community developed Plumed[15,16] library patched with LAMMPS[7]. In WTM simulations, a bias potential is added along the collective variables (CVs), to probe the free-energy landscape along the CVs. In this work we used two dihedral angles (described in main manuscript) as CVs for WTM simulations. In WTM simulations, we used gaussian width of 0.1 radian, initial gaussian height of 0.05 eV, biasfactor of 12, grid spacing of 0.05 radian, and bias deposition rate of 100.

To maintain stability of WTM simulations, we applied upper walls on four Zn-O distances (near to the considered phenylene ring) at a value of 2.4 Å with force constant of 25 eV-Å$^2$. In addition to these, upper and lower walls were applied to two $O_2$-Zn-Zn-$O_2$ dihedral angles) at a value of 0.85 radian with force constant of 25 eV-radian$^2$. We performed WTM simulations at different temperatures in isothermal ensemble.

```
Δ_cl = 0.0; Δ_ca = 0.0; Δ_b = 0.0; Δ_a = 0.0; Δ_d = 0.0;

Cr = [[],[],[],[],[],[]]; # for a, b, c, α, β, γ
Br = [[],[],[],…,[]];    # for l bond types
Ar = [[],[],[],…,[]];    # for m angle types
Dr = [[],[],[],…,[]];    # for n dihedral types

for i = 1 to N_C   # N_C number of configurations
{   check = 0;
    S = get_CBAD(i); #This function will return CBAD values of a configuration

for j = 1 to 3    # for a, b, c
   { if(int(S.C[j]/Δ_cl) not in Cr[j] ) { check = check + 1; } }
for j = 4 to 6    # for α, β, γ
   { if( int(S.C[j]/Δ_ca) not in Cr[j] ) { check = check + 1; } }

for j = 1 to l    # l bond types
  { for k = 1 to B_j    # B_j number of bonds of type j
     { if(int(S.b_k^j/Δ_b) not in Br[j] ) { check = check + 1; } } }

for j = 1 to m    # m angle types
  { for k = 1 to A_j    # A_j number of angles of type j
     { if(int(S.a_k^j/Δ_a) not in Ar[j] ) { check = check + 1; } } }

for j = 1 to n    # n dihedral types
  { for k = 1 to D_j    # D_j number of dihedrals of type j
     { if(int(S.d_k^j/Δ_d) not in Dr[j] ) { check = check + 1; } } }

    if(check== 6+nbonds + angles + dihedrals)
    { include_in_training_set(S);
      update(Cr,Br,Ar,Dr,S);
    }
}
```

**Figure S1:** A pseudocode illustrating CBAD active learning algorithm for deciding which configuration to include in the training set. This algorithm compares cell-parameters, bonds, angles, and dihedrals (CBAD) of a given configuration with the existing values of the training set and includes the configuration in training set if either of the values is not included in the training set.

(a) Training

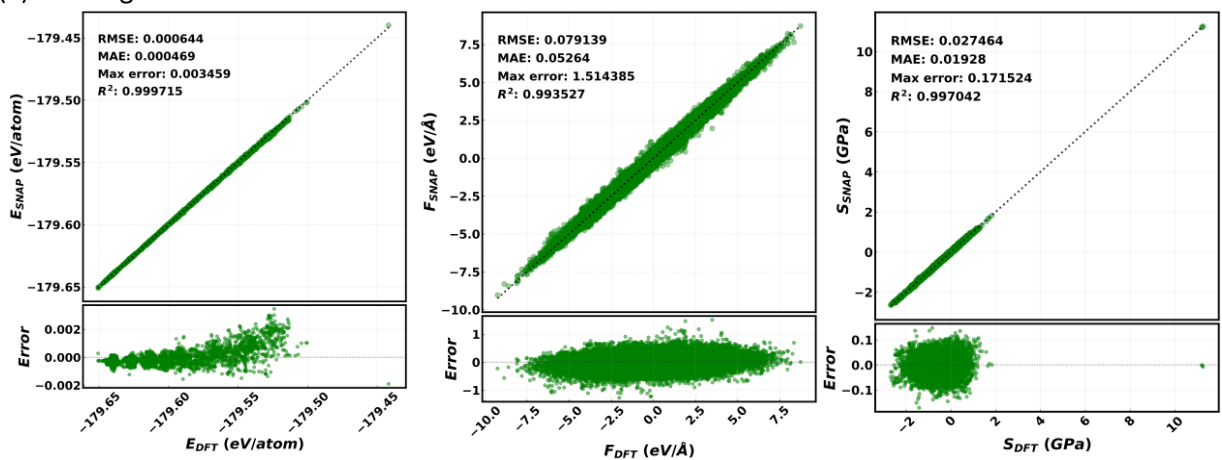

(b) Test A (AIMD configurations)

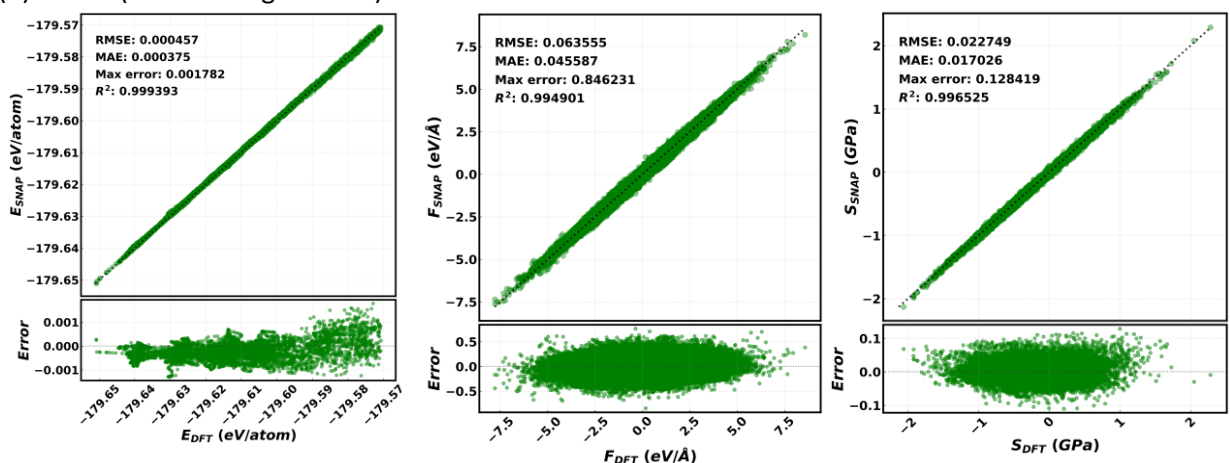

(c) Test B (MD configurations)

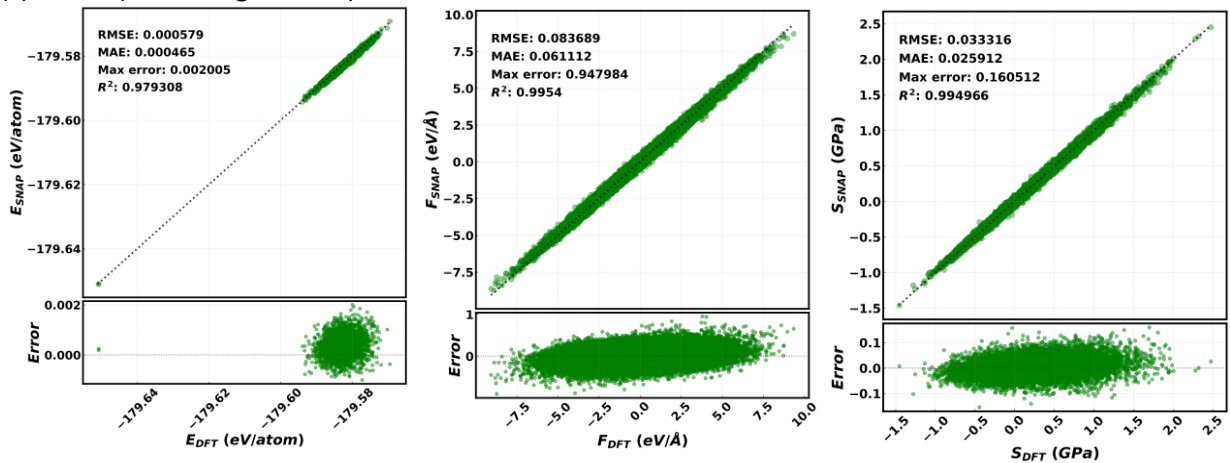

**Figure S2:** Parity plots of energy (E), forces (F), and stress (S) values for (a) training set (3138 configurations), (b) test set A (4956 configurations), and (c) test set B (2379 configurations) of ZIF-8. These parity plots compare EFS values of different configurations of ZIF-8, obtained using SNAP (trained over 3138 configurations) and DFT.

(a) Training

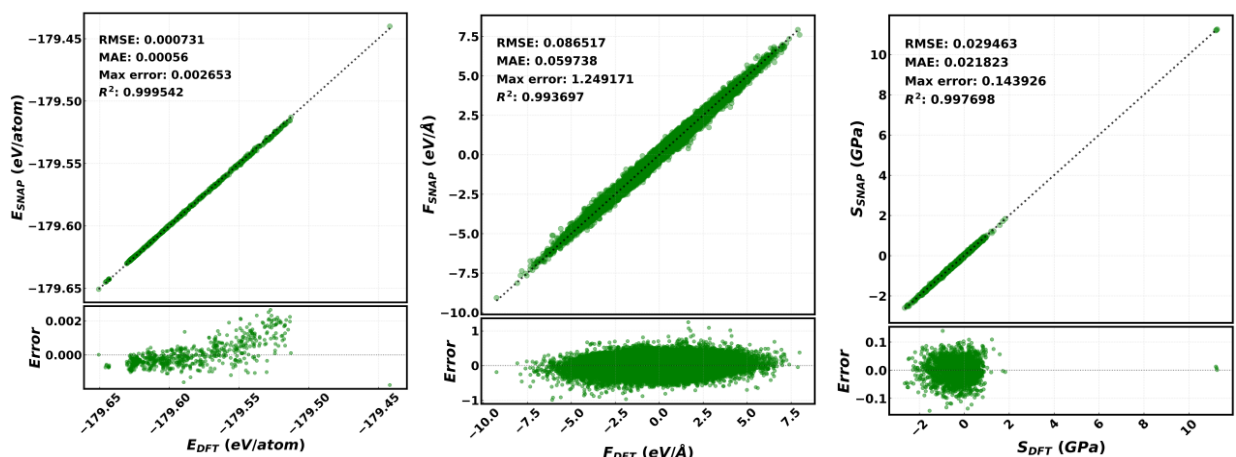

(b) Test A (AIMD configurations)

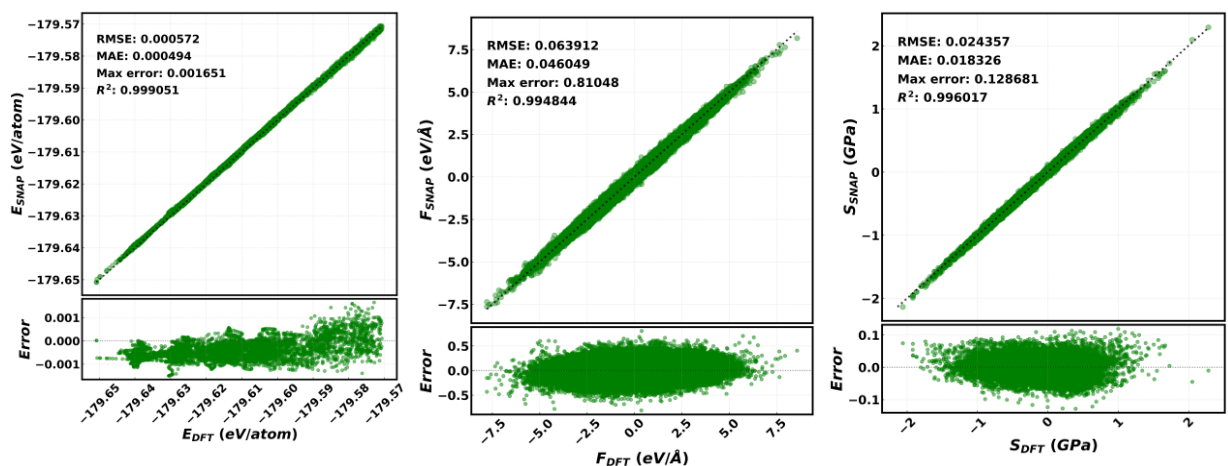

(c) Test B (MD configurations)

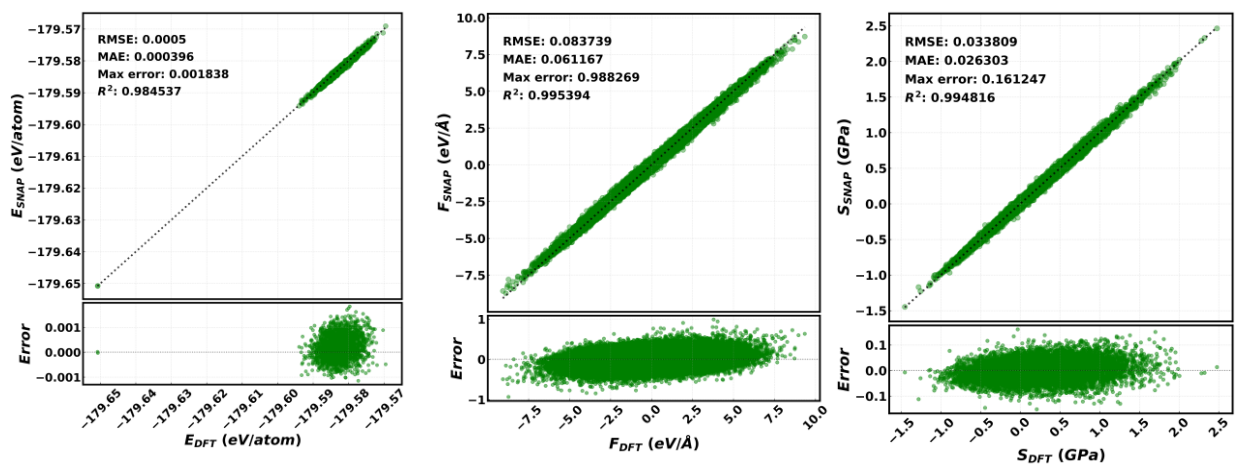

**Figure S3:** Parity plots of energy (E), forces (F), and stress (S) values for (a) training set (672 configurations), (b) test set A (4956 configurations), and (c) test set B (2379 configurations) of ZIF-8. These parity plots compare EFS values of different configurations of ZIF-8, obtained using SNAP (trained over 672 configurations) and DFT.

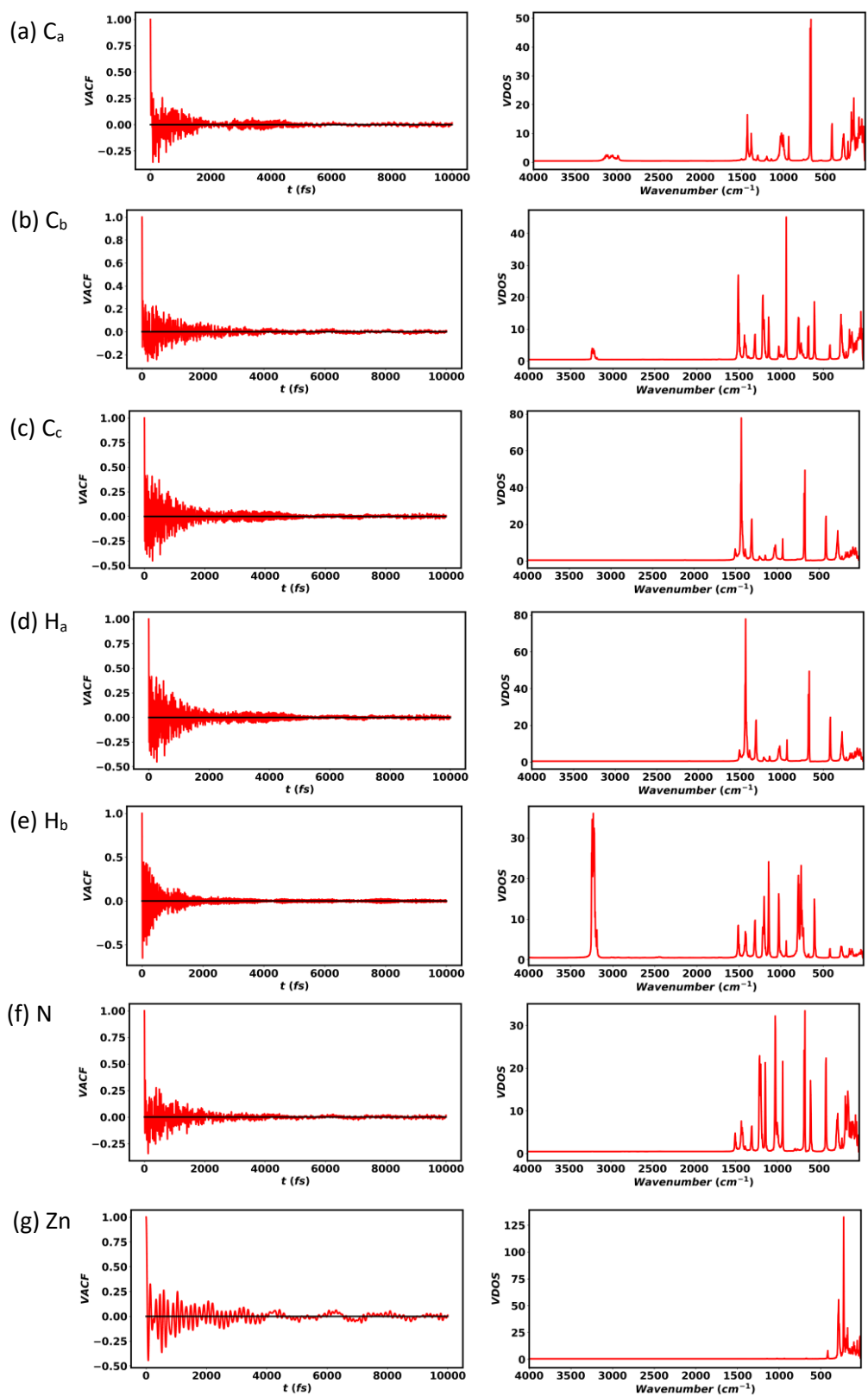

**Figure S4:** Velocity autocorrelation function (left) and partial vibrational density of states (right) for each atom type of ZIF-8.

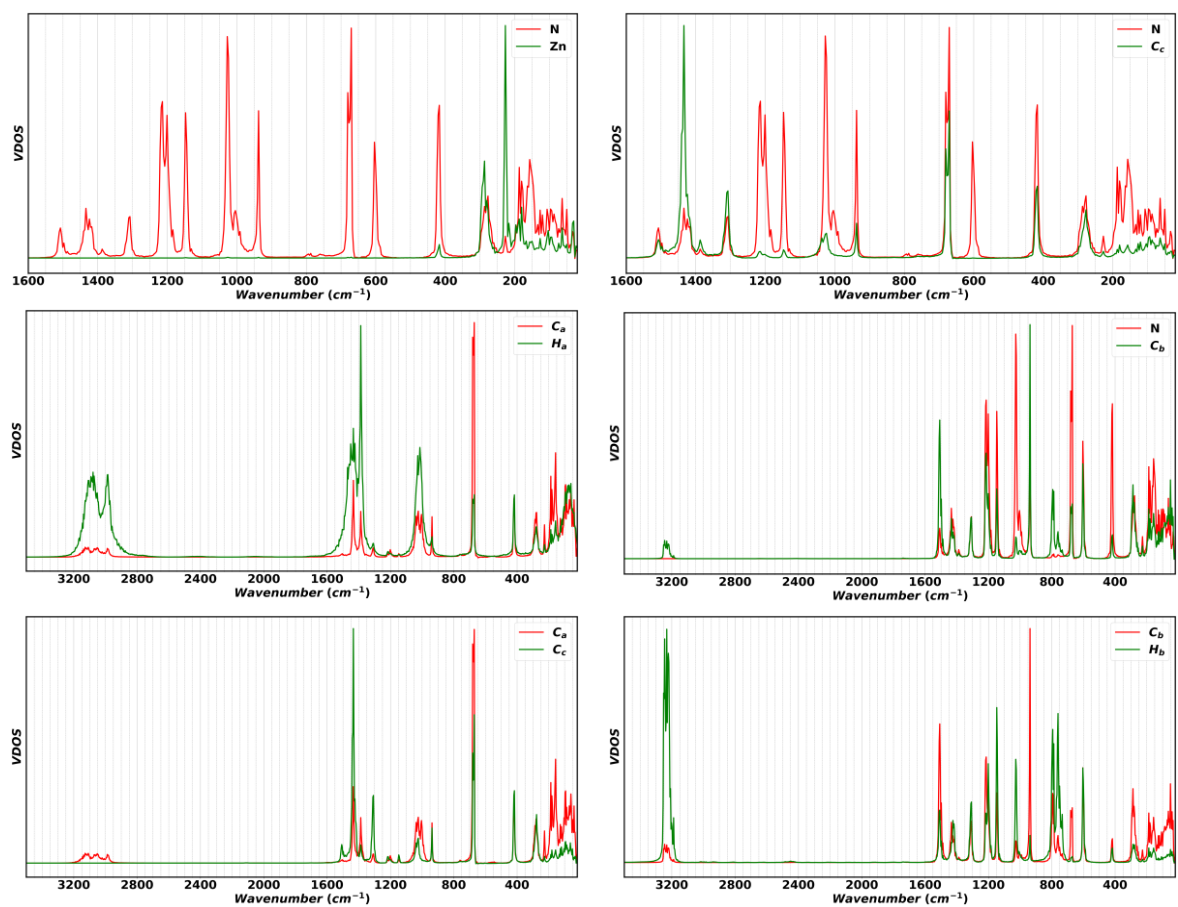

**Figure S5:** Comparison of normalized partial vibrational density of states between neighbouring atoms of ZIF-8.

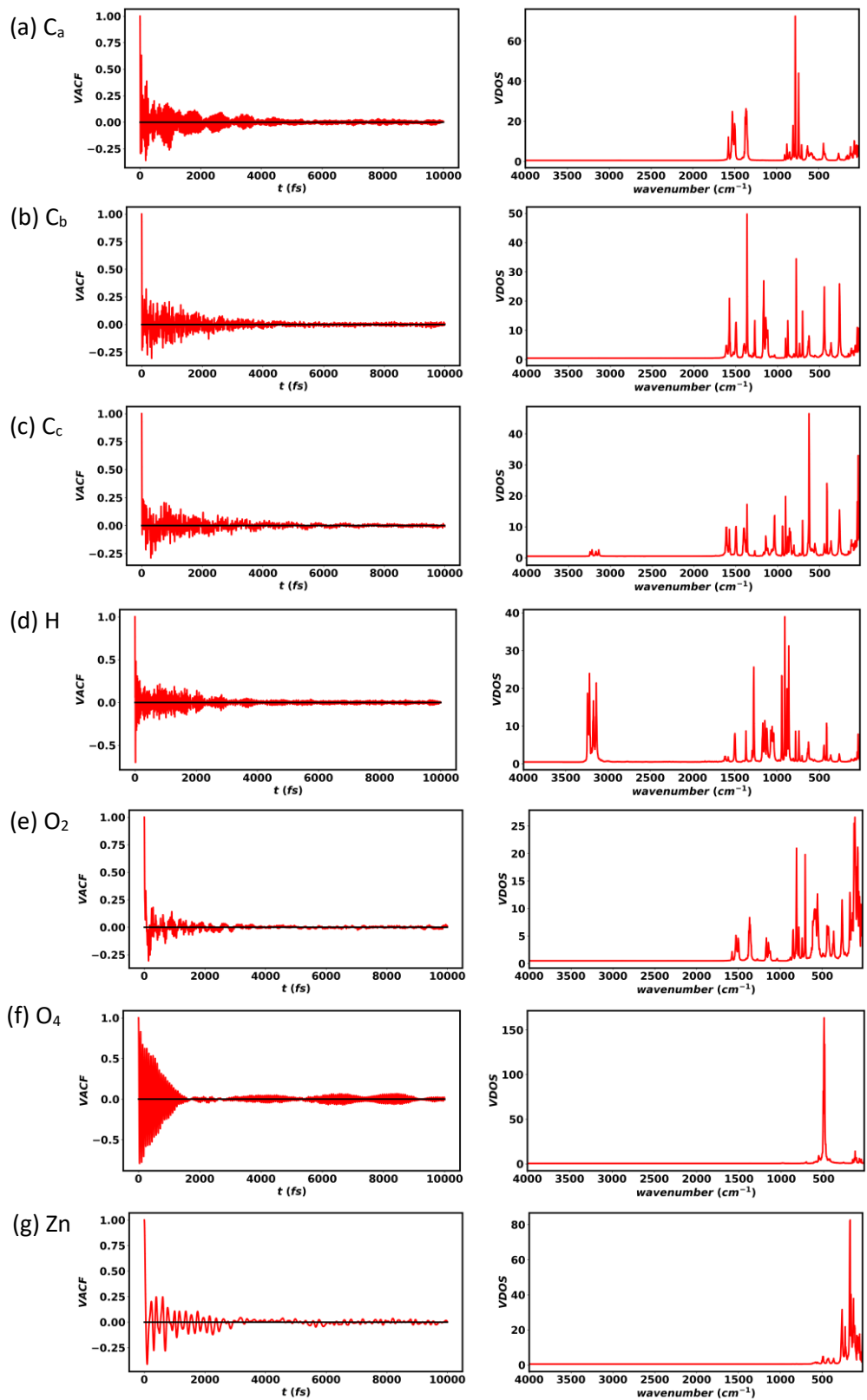

**Figure S6:** Velocity autocorrelation function (left) and partial vibrational density of states (right) for each atom type of MOF-5.

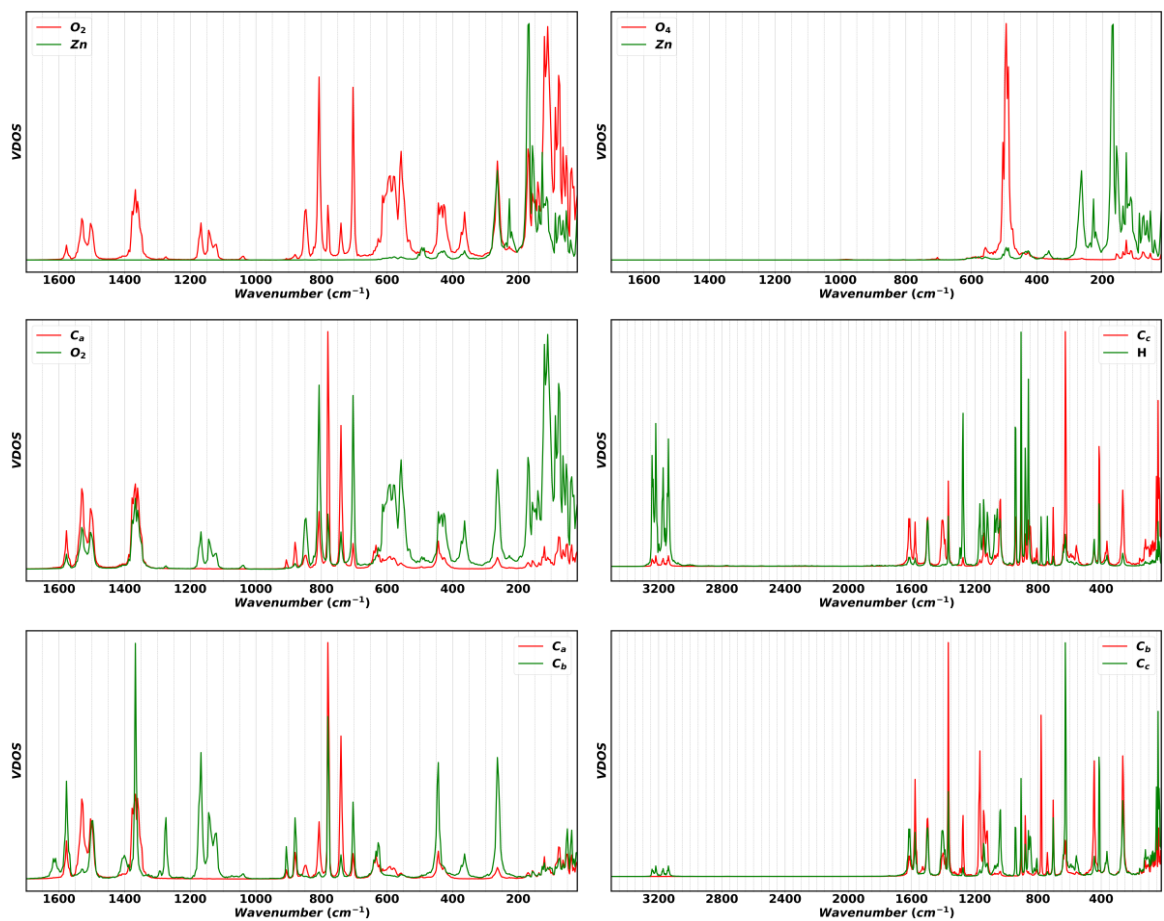

**Figure S7:** Comparison of normalized partial vibrational density of states between neighbouring atoms of MOF-5.

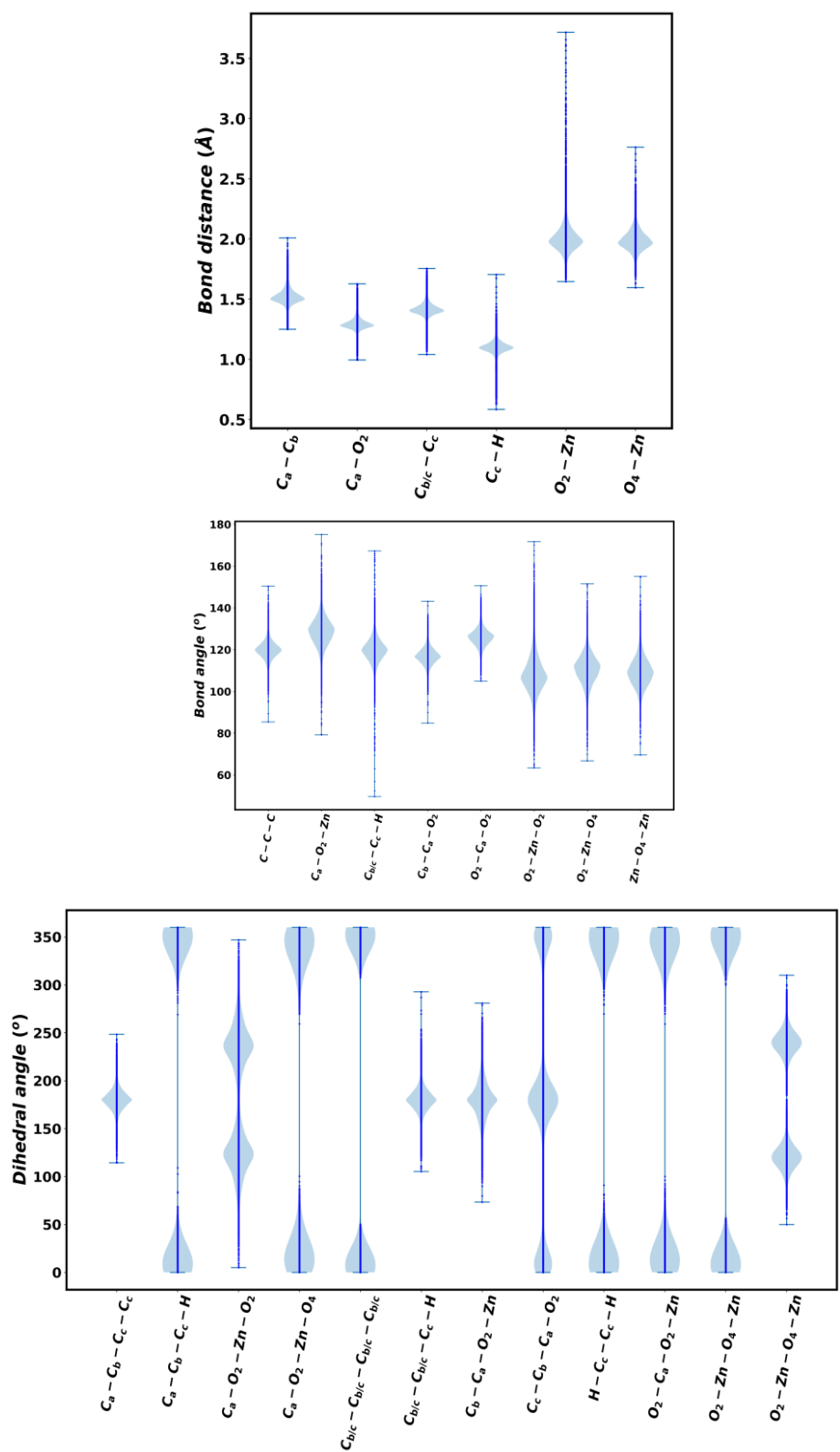

**Figure S8:** Distribution of bond distance, bond angles, and dihedral angles of MOF-5 training set configurations.

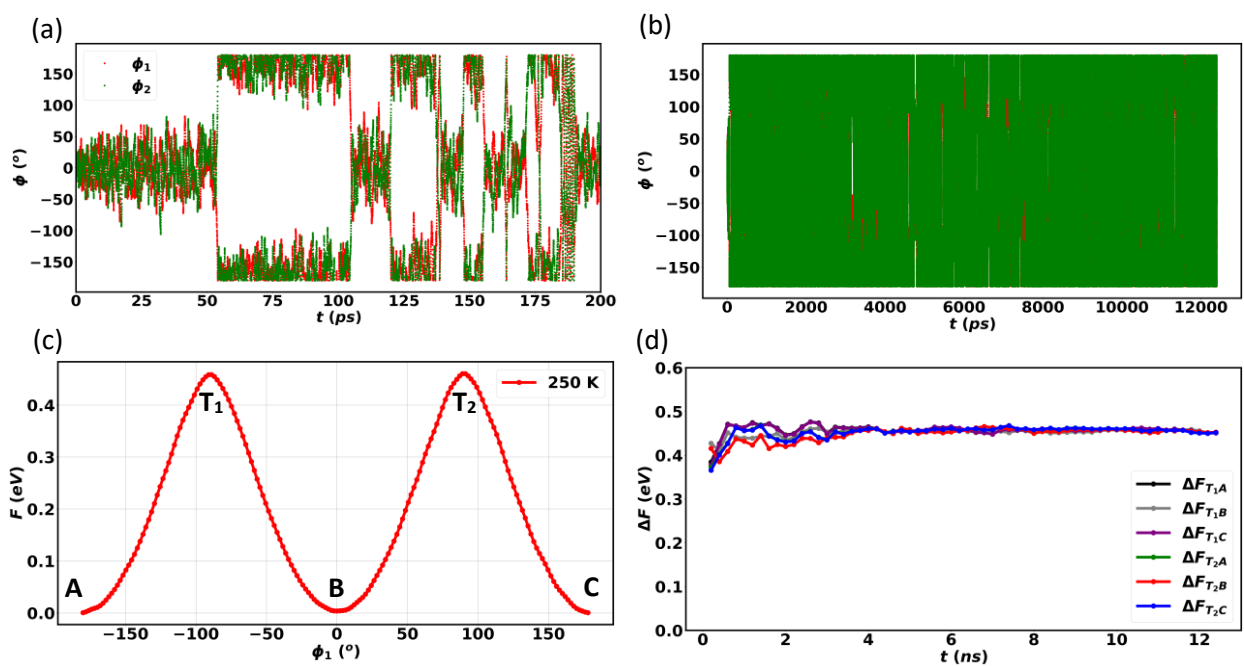

**Figure S9:** Distribution of dihedral angle collective variables as a function of simulation time of (a) 200 ps and (b) 12000 ps. (c) Free energy surface as a function of one collective variable. **A**, **B**, and **C** denotes stable states and **T₁** and **T₂** denotes transition states. (d) Evolution of rotational barrier heights (free energy difference between stable and transition states) as a function of simulation time.